\documentstyle[aps,preprint,tighten,epsfig,amssymb]{revtex}

\begin{document}
\draft

\title{
Probing in-medium vector meson decays by
double-differential di-electron spectra
in heavy-ion collisions at SIS energies 
}

\author{Gy. Wolf,$^{a,b}$
O.P. Pavlenko,$^{a,c}$
B. K\"ampfer,$^a$
}

\address{
$^a$Forschungszentrum Rossendorf, PF 510119, 01314 Dresden,
Germany\\
$^b$ KFKI RMKI, H-1525 Budapest, POB 49, Hungary\\
$^c$Institute of Theoretical Physics,
252143 Kiev, - 143, Ukraine
}

\maketitle

\begin{abstract}
Within a transport code simulation for heavy-ion collisions
at bombarding energies around 1 AGeV, we demonstrate that
double-differential di-electron spectra with suitable
kinematical cuts are useful to isolate
(i) the $\rho$ meson peak even in case of strong broadening, and
(ii) the in-medium $\omega$ decay contribution.
The expected in-medium modifications of 
the vector meson spectral densities can thus be probed
in this energy range via the di-electron channel.
\\[3mm]
{\it PACS number(s):} 25.75.+r, 14.60.-z, 14.60.Cd\\[3mm]
\end{abstract}


\section{Introduction} 

One of the main goals of the experiments with the HADES
detector \cite{Friese} at the heavy-ion synchrotron 
SIS18 at GSI/Darmstadt
is the search for a direct evidence of in-medium modifications
of hadrons via the di-electron ($e^+ e^-$) channel.
This goal is supported by various theoretical indications
concerning an important sensitivity of vector mesons
to partial restoration of chiral symmetry in a dense and hot
nuclear medium \cite{Rapp_Wambach}. 
In particular, as shown within lattice QCD
\cite{Karsch}, the chiral quark condensate 
$\langle \bar q q \rangle$ as order parameter of the
chiral symmetry breaking decreases with increasing baryon
density and temperature. Since $\langle \bar q q \rangle$ is
coupled to the vector meson spectral densities one can expect
considerable in-medium effects of light vector mesons
which manifest themselves, e.g., as resonance mass shifts
and broadening of widths \cite{Brown}.

In spite of the efforts spent to predictions
\cite{Hatsuda,Weise,Leupolt,Leinweber,Hofmann,Gutsche,Schneider,Bengt,we},
there is no commonly accepted and reliably quantitative estimate
of the change of the mentioned parameters or spectral functions
in a nuclear medium. For instance, within the QCD sum rule analysis
\cite{Hatsuda,Weise,Leupolt} the in-medium modification of the vector
meson masses in zero-width approximation is dictated by the density
and temperature dependence of the quark condensates.
While the two-quark condensate $\langle \bar q q \rangle$ can be
estimated by a model independent approach with reasonable accuracy,
the so far poorly known four-quark condensates like  
$\langle (\bar q q)^2 \rangle$ lead to uncertainties
for the prediction of the vector meson mass shift \cite{PLB}.
From this point of view, direct measurements of the vector meson
masses and widths, as being at reach with the high-resolution
HADES detector system \cite{Friese}, can also shed light on
fundamental problems such as the in-medium change of the four-quark
condensate.

Since the most pronounced vector meson mass shift is expected for
large baryon density in the nuclear medium, there is the hope
to verify the effect in medium-energy heavy-ion collisions at SIS18
energies,
where baryon densities up to $3 n_0$ can be reached according to 
model calculations \cite{Cass}. 
Unfortunately, in the case of heavy-ion collisions such measurements
face a number of problems related to the expansion dynamics 
and the spatial inhomogeneity of the produced
dense matter. Due to the collective expansion, the baryon density
$n$ depends on time so that obviously a time average of the
mass shift is to be expected, i.e. the mass shift effect is distributed
over an interval related to the temporal change of $n$. 
Also spatial variations
of $n$ will make the situation more complicated.

To extract the wanted information on the in-medium meson spectral
function in heavy-ion collisions one has also to take into account
that the di-electron spectrum $dN/dM^2$ as a function of the di-electron
invariant mass $M$ is actually a convolution of the considered
spectral function and the probability to create the
vector meson with given energy and momentum.
The latter one usually drops down very fast with the energy
making the resulting di-electron spectrum very steep.
This can make the identification of the possible mass shift,
superimposed to collision broadening and further in-medium
broadening, in particular for the $\rho$ meson,
rather difficult. An additional problem arises for the
$\omega$ meson because of its comparatively long life time
which, however, may change in a dense medium \cite{Schneider}.
To measure an in-medium mass shift of the $\omega$ resonance one
should obviously separate lepton pairs produced by decays in the dense
medium from such ones which originate from decays after the 
disintegration of the dense system.

The aim of the present note is to explore the extension of the usual
analysis of the invariant mass spectrum $dN/dM^2$ to the more
informative double-differential spectrum
$dN/dM^2 \, dM_\perp^2$, where 
$M_\perp = \sqrt{M^2 + \vec Q_\perp^2}$ 
is the transverse mass of the pair with transverse two-momentum
$\vec Q_\perp$ perpendicular to the beam axis.
As shown in \cite{our} within a schematic fireball model, such an
extension with suitable kinematical cuts
allows to avoid the majority of the above mentioned
difficulties in extracting information on 
the in-medium vector meson spectral function.
Here we present results of our calculations
of double-differential di-electron spectra in heavy-ion collisions
at SIS18 energies obtained within a transport
model of Boltzmann-Uehling-Uhlenbeck (BUU) type. 

\section{Di-electron production and vector meson spectral function} 

As elementary processes for the $e^+ e^-$ production we employ 
pion annihilation $\pi^+ \pi^- \to \rho \to e^+ e^-$,
direct vector meson decays $V \to e^+ e^-$ 
with $V = \rho, \omega$,
pion - nucleon and nucleon - nucleon bremsstrahlung,
and Dalitz decays of mesons, $\Delta$ and $N^*$.
Pions are produced solely in baryonic resonance decays, and
these resonances are created either in baryon - baryon collisions via
$B B \to B B$ with $B = N, \Delta, N^*$ or in meson - baryon
collisions $M B \to B$ with $M = \pi, \rho, \sigma$,
while the ''direct vector mesons'' emerge in the elementary
reactions $B B \to V N N$ or $\pi B \to V N$.
Our treatment of these processes is described in some detail in
\cite{Gyuri}. 
To be specific, let us list our perturbative 
treatment of the $\rho$ and $\omega$ channels.

We use the parameterization \cite{Gale_Kapusta} of the total cross section
$\sigma^{\rm vac}$ for $\pi^+ \pi^- \to \rho \to e^+ e^-$ in vacuum
\begin{equation}\label{eq.1}
\sigma^{\rm vac} (M) = \frac{4 \pi}{3} \left( \frac{\alpha}{M} \right)^2
\sqrt{1 - \frac{4m_\pi^2}{M^2}} \, 
\frac{\tilde m_\rho^4}
{(M^2 - \hat m_\rho^2)^2 + (\tilde m_\rho \tilde \Gamma_\rho)^2}
\end{equation}
with $\tilde m_\rho = 775$ MeV, $\hat m_\rho = 761$ MeV,
$\tilde \Gamma_\rho = 118$ MeV
to generate the in-medium spectrum by 
\begin{equation}\label{eq.5}
\sigma (M) = \sigma^{\rm vac} (M) \,
\frac{m_\rho^2 (n) \, A(M; m_\rho (n), \Gamma_\rho (n))}
{m_\rho^2 (n = 0) \, A(M; m_\rho (n = 0), \Gamma_\rho (n = 0))}, 
\end{equation}
where the spectral function 
is
\begin{equation}\label{eq.6}
A (M; m_\rho, \Gamma_\rho) = \frac1\pi 
\frac{m_\rho \Gamma_\rho}{(M^2 - m_\rho^2)^2
+ (m_\rho \Gamma_\rho)^2} .
\end{equation}
with mass dependent width 
\begin{equation}
\Gamma_\rho (n) = \Gamma_\rho^{(0)} (n) \,
\left( \frac{\sqrt{M^2 - 4 m_\pi^2}}
{\sqrt{m_\rho^2 (n) - 4 m_\pi^2}} \right)^3.
\end{equation}
Below we employ certain parameterizations of 
$\Gamma_\rho^{(0)} (n)$ and $m_\rho (n)$.

Similar to the case of $\pi^+ \pi^-$ annihilation, the distribution
of in-medium decays of directly produced $\rho$ mesons into
$e^+ e^-$ follows in local density and pole mass approximation 
from\footnote{
This contribution is of sub-leading order (see below),
therefore, an estimate of the maximum yield at pole position
is sufficient here.}
\begin{equation}
\frac{d \sigma_{e^+ e^-}}{d M} = 
\sigma_\rho(\sqrt{s}) \, 2 M \,
A(M; m_\rho (n), \Gamma_\rho (n)) \,
Br(\rho \to e^+ e^-)
\end{equation}
with the branching ratio of 
$Br(\rho \to e^+ e^-) 
= \Gamma (\rho \to e^+ e^-) / \Gamma (\rho, {\rm tot})$,
so that all in-medium effects are related
to the modification of the vector-meson
spectral function. 
We take the default value of $Br$ as in vacuum, i.e., assume that
the e.m.\ decay width and the total width change in the
same manner.
Such a choice of the branching ratio
allows us to avoid an artificial change of the
relative contributions of $\rho$ and $\omega$
mesons (see below for the case of the density dependent
$\omega$ mass shift).  
In Eq.~(5), $\sigma_\rho (\sqrt{ s })$ 
is a parameterization of the total cross section.
For $ NN \to NN \rho$ we use
$\sigma_\rho = F_\rho \, S \, \frac{x^y}{(\xi + x)^z}, $
where $x = (\sqrt{s} - m_\rho (n)) / {\rm GeV}$ and
$S = 1$, $F_\rho = 1.5$ mb, $\xi = 1.4$, $y = 1$, z = 2. In case of
$N \Delta$ ($\Delta \Delta$) collisions 
the scaling factor $S = 0.75$ (0.5) is employed.
The $\pi N \to N \rho$ reaction is similarly parameterized, but
$F_\rho = 1.5$ mb, $\xi = 0.018$, $y = 2.2$, $z = 3.5$.
The scaling factors $S$ here are  
0.375 for $\pi \Delta^{++} [\Delta^-]$, 
0.125 for $\pi^+ \Delta^0$, $\pi^- \Delta^+$, and
0.25 for $\pi^0 \Delta$.

The $e^+ e^-$ decay spectrum of $\omega$ mesons follows from 
\begin{equation}
\frac{d \sigma_{e^+ e^-}}{d M} = \sigma_\omega
\int_0^\infty dt \, 
\exp \left\{ - \Gamma_\omega (n) \gamma t \right\}  \,  
2 M \,
A(M; m_\omega (n), \Gamma_\omega (n)) \,
\Gamma_\omega (n) \,
Br(\omega \to e^+ e^-),
\label{eq.7}
\end{equation}
where the path
through matter with local density $n(t, \vec x)$ is followed.
$\gamma$ is the Lorentz factor according  to the velocity, and
$Br(\omega \to e^+ e^-)$ denotes the vacuum branching ratio.
The production cross section $\sigma_\omega$
is given by the above parameterization
with $F_\omega = 0.36$ mb, $\xi = 1.25$, $y = 1.4$, $z = 2$
for the reaction $NN \to NN \omega$
with the same scaling factors as for the $\pi N \to N \rho$
recation and corresponding isospin channels.
The reaction $\pi N \to N \omega$ follows from
$F_\omega = 1.38$ mb, $\xi = 0.0011$, $y = 1.6$, $z = 1.7$,
where the appropriate scaling factors $S$ are
0.5 for $\pi^0 N [N*]$,
1 for $\pi^ + n$ or $\pi^- p$,
and all others as above.

\section{Numerical results} 

According to our experience \cite{Gyuri}, the di-electron spectra
roughly scale when changing the mass numbers
of the colliding nuclei. We choose, therefore, the scaled cross section
$d \sigma^{\rm scal} / dM = \pi R^2 A^{-2} \,
dN(b=0)/dM$ with $R = A^{1/3} \, 1.124$ fm.
Since we are going to demonstrate the
advantage of certain observables for the case of strong
medium effects, we focus on central collisions
of gold nuclei at a beam energy of 1 AGeV.

First we use the parameterization
$\Gamma^{(0)}_\rho (n)  = a \tilde \Gamma_\rho$,
$m_\rho (n) = \tilde m_\rho - \Delta m_\rho$
with constant values of $a$ and $\Delta m_\rho$.
This is schematic and aimed at verifying the previous
finding \cite{our} that a strong in-medium broadening of the
vector meson widths smears out the $\rho$ meson peak in the
invariant mass spectrum, while the double-differential spectrum
still allows to identify the $\rho$ peak. This set should also
illustrate the separate effects of mass shifts. 
While in \cite{our} a fireball model with thermal and chemical
equilibrium was used to calculate the di-electron spectra,
here both these conditions are not longer required. 
Moreover, to contrast the present study
with the one in \cite{our} we switch off any rescattering of the
$\rho$ and $\omega$ mesons once created thus not allowing
thermalization and chemical equilibration of these species. 

In Fig.~1 we exhibit the resulting double-differential cross section
$d\sigma^{\rm scal} / dM \, dM_\perp$ 
for the parameters
$a = 3$ and $\Delta m_\rho = 0$. Let us first consider the invariant mass
spectrum arising by the full $M_\perp$ integration
(see lower panel in Fig.~1). We are interested
in the invariant mass region $M > 400$ MeV, where obviously
the pion annihilation channel $\pi^+ \pi^- \to \rho \to e^+ e^-$
dominates over Dalitz decay and bremsstrahlung contributions.  
The $\omega$ peak sticks out, but the chosen broadening
is so strong that any peak structure, attributable to the
$\rho$ meson by pion annihilation, disappears. Instead,
a knee is reminiscent of the $\rho$.
This is the general expectation that a spectroscopy of the
$\rho$ meson becomes impossible due to broadening \cite{Weise}.
However, selecting an appropriate $M_\perp$ window, say
850 MeV $< M_\perp <$ 950 MeV, the $\rho$ peak becomes clearly
visible, even for the assumed extreme broadening 
(see upper panel in Fig.~1).
Due to phase space weighting, the $\rho$ peak position appears
slightly shifted to smaller mass. At the high-$M$ wing the direct
$\rho$ contribution and the pion annihilation yield compete.
The $M_\perp$ window around
900 MeV seems most suitable, as it represents a compromise
of suppressing unwanted background, not cutting off kinematically
the $\rho$ peak, and still allowing for sufficient statistics. 

Next let us consider the effect of a mass shift, which is strongest
for the $\rho$ meson according to many predictions. 
Fig.~2 demonstrates for $a = 3$, $\Delta m_\rho = 150$ MeV
that the above found features are persistent 
also in case of quite drastic mass shifts. 
The upper panel in Fig.~2 illustrates, indeed, 
that such a strong mass shift of the $\rho$ meson
is clearly visible even in case of the strong broadening
in the spectrum 
$d \sigma^{\rm scal} / d M \, dM_\perp 
\vert_{M_\perp = 850 \cdots 950 {\rm MeV}}$,
while for the chosen parameters the invariant mass spectrum
does not exhibit any $\rho$ peak, due to the assumed
strong broadening (see lower panel in Fig.~2).

Figs.~1 and 2 show that for identifying the $\omega$ meson still 
the invariant mass spectrum is enough: For in-medium changes covered
by the parameterization of set (i) there is no obvious advantage 
by a particular $M_\perp$ selection. The important point here is that
the $\omega$ meson rides on the pion annihilation
background. 

Note that in Figs.~1 and 2 the distributions are according
to the spectral function Eq.~(\ref{eq.6}) and the phase space; in line with
the model assumptions (no density dependence of $\Delta m_V$
and $\Gamma_V$, no rescattering), there is no need of an explicit
propagation of the spectral function.  

Let us now turn to
temporal and spatial variations of the density which cause an
additional distribution of the vector meson decay strength. 
Also here, the usefulness of the double-differential cross section 
with appropriate kinematical cuts still applies.
We focus on an illustration
of the $\omega$ mass shift. Since the $\omega$ in vacuum is a narrow
resonance, often it is assumed to survive the dense stages during
a heavy-ion collision and to decay after freeze-out with its vacuum
parameters. Therefore, the $\omega$ in-medium spectroscopy is
considered as more useful in reactions of elementary projectiles
with a nucleus under special kinematical conditions in which
the $\omega$ is roughly at rest and decays inside the static matter.
We would like to analyze here the prospects of the $\omega$
in-medium spectroscopy in heavy-ion collisions.

In focusing on the $\omega$ meson, we leave for the moment being the 
$\rho$ meson parameters unattached 
(i.e., take vacuum parameters)
and choose 
$m_\omega (n) = m_\omega - \delta m_\omega \, n / n_0$, 
$\Gamma_\omega (n) = \Gamma_\omega (n = 0) + 
\delta \Gamma_\omega n / n_0$
with $\delta m_\omega = 70$ MeV and
$\delta \Gamma_\omega = 50$ MeV entering the spectral
function of the $\omega$ which is of the form of Eq.~(3). 
Given the still sufficiently narrow width of the in-medium $\omega$
we propagate the peak according to local conditions
and distribute it at each time instant with the spectral
function Eq.~(\ref{eq.6}) and sum up the decay products,
cf.\ Eq.~(\ref{eq.7}).
These procedure
is acceptable only for sufficiently narrow resonances
(see discussion in \cite{Knoll}). 
Since the number of $\omega$ mesons is
normalized, an increase of the width means a reduction of the peak
height. 
For the selected parameters,
the $\omega$ peak for the above in-medium parameters has a
contribution from in-medium decays and vacuum decays. The first
one shows clearly the down shift and the smearing due to the
density dependence, see lower panel in Fig.~3. 
The vacuum peak sticks not longer so much above the pion annihilation yield.
The in-medium and vacuum $\omega$ contributions 
compete with the $\rho$ contribution from $\pi^+ \pi^-$ annihilation.
To increase the relative contribution of in-medium $\omega$
decays, in \cite{Cassing} it has been proposed to select low-$Q_\perp$
di-electrons which stem from very slow $\omega$ mesons which decay
in the medium, in contrast to fast $\omega$ mesons which traverse
quickly the medium and decay outside. This feature is also
seen in our simulations for heavy-ion collisions
in the upper panel in Fig.~3, where we selected di-electrons with
$Q_\perp < 100$ MeV: the vacuum contribution becomes
significantly less, while the in-medium $\omega$ contribution
is clearly seen as a shifted bump riding on the
$\pi^+ \pi^-$ annihilation background. This is in contrast to 
the invariant mass spectrum which is integrated over all $Q_\perp$
(see lower panel in Fig.~3) where the strong competition
between the wanted $\omega$ meson signal and the
$\pi^+ \pi^-$ annihilation background makes a conclusion on
a in-medium $\omega$ mass shift difficult. Therefore, the
measurement of the double-differential di-electron spectrum at low
values of $Q_\perp$ offers much better chances to verify a
$\omega$ meson mass shift.
Further cuts, e.g. setting an additional narrow repidity window
around mid-rapidity, suppress too strongly the
yield but do not enhance efficiently enough the in-medium 
contribution.

With respect to the latest HADES $e^+ e^-$ measurement
in carbon - carbon collisions, we mention that
in our simulations of central collisions C + C
the $Q_\perp$ dependence is similar to the case of Au + Au.
However, for C + C the ''in-medium $\omega$ shoulder''
appears below the $\pi^+ \pi^-$ channel,
since the in-medium $\omega$ contribution is attenuated due to
the shorter life time and the smaller volume of compressed 
nuclear matter. 
The vacuum decay contribution dominates, even for small values of
$Q_\perp$.

Finally, we would like to comment on the above choice of
$Br = Br^{\rm vac}$, which assumes that $\Gamma (\omega \to e^+ e^-)$
and $\Gamma (\omega, \rm tot)$ scale in the same manner in medium.
In contrast, often an in-medium change of  $\Gamma (\omega, \rm tot)$
is assumed, and $\Gamma (\omega \to e^+ e^-)$ is kept constant
(cf.\ \cite{Cass} for a discussion). In the latter case
the in-medium contribution is noticeably suppressed,
as evidenced in Fig.~4. Clearly, this issue, as the
treatment of dynamical in-medium effects of $\rho$ and $\omega$
on a common footing, needs further investigations.

\section{Summary} 

Our results can be summarized as follows: 
(i) The spectrum $d\sigma /dM \, dM_\perp$ for 
850 MeV $< M_\perp <$ 950 MeV 
as a function of $M$ is dominated in the region
400 MeV $< M <$ 800 MeV by the channel 
$\pi^+ \pi^- \to \rho \to e^+ e^-$
with a pronounced $\rho$ bump. This gives a good chance to measure the
in-medium $\rho$ meson spectral function
by a double-differential di-electron spectrum
in heavy-ion collisions at SIS18 energies. 
(ii) By selecting di-electrons with small enough transverse momentum
$Q_\perp < 100$ MeV one can isolate the in-medium
$\omega$ decay contribution which signals
the in-medium mass shift and a broadening of the $\omega$
meson. This $Q_\perp$ dependence may be useful for
future $\omega$ spectroscopy. 
(iii) Bremsstrahlung and Dalitz decays are small in the
region $M >$ 400 MeV, for the considered beam energies.

Our findings, which are not to be understood as quantitative
predictions, 
also show that fairly independently of the specific form
of the vector meson spectral function the double-differential 
di-electron spectrum with suitable cuts allows a better
access to the in-medium  effects and, therefore, can be an
appropriate measurement strategy in HADES experiments.

{\bf Acknowledgments:}
Useful discussions with 
H.W. Barz, W. Cassing, and G. Zinovjev are acknowledged.
The work is supported by BMBF 06DR 921,
DAAD Hungarian - German exchange programme 324-PPP,
and the National Fund for Scientific Research of Hungary
OTKA T32038 and T30855. Gy.W. and O.P.P. thank for the warm
hospitality of the nuclear theory group in the Research Center
Rossendorf and support by STCU 15a, CERN-INTAS 2000-349,
and NATO-2000-PST.CLG977482.

\renewcommand{\baselinestretch}{1.2}
\renewcommand{\baselinestretch}{1.5}

\begin{figure} 
\centerline{
\epsfig{file=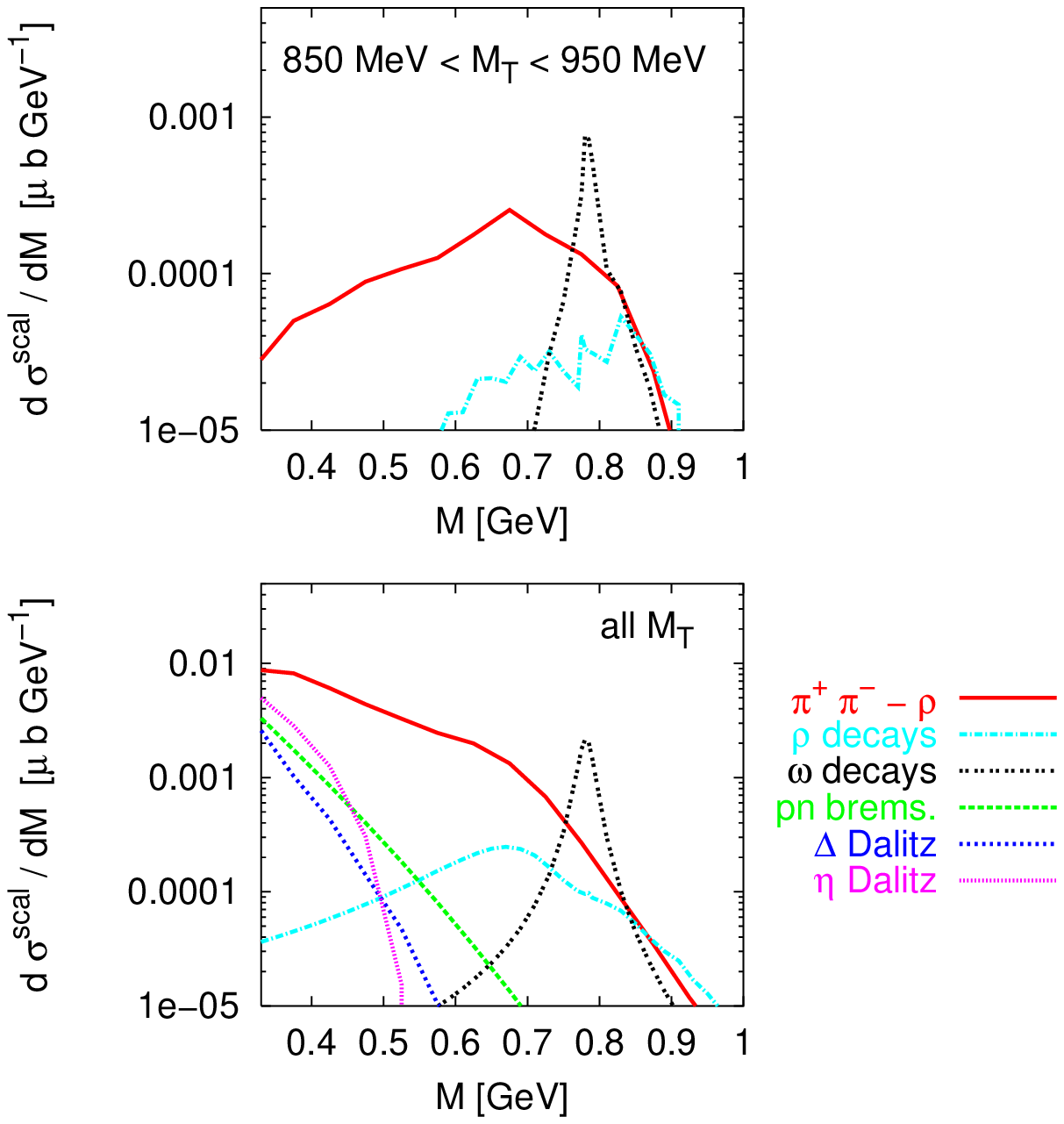,height=15cm}
}
\vskip 9mm
\caption{Contributions to the di-electron spectrum 
$d \sigma^{\rm scal} / d M$ for various bins of $M_\perp$
for central collisions Au(1 AGeV) + Au.
A resonance broadening by a factor $a = 3$ is
assumed as in \protect\cite{our} for both $\rho$ and $\omega$; 
no mass shift.}
\label{fig.1}
\end{figure}

\begin{figure} 
\centerline{
\epsfig{file=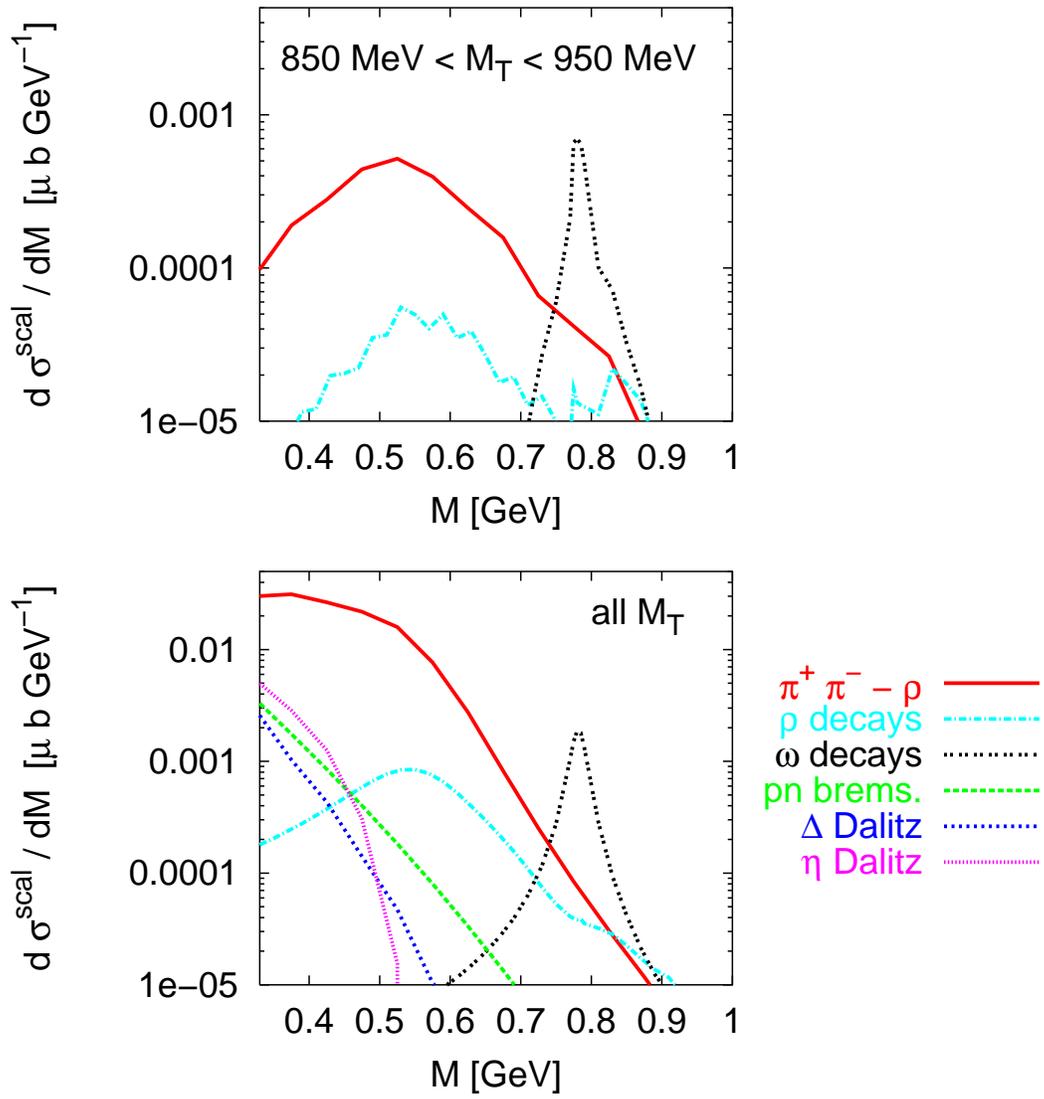,height=15cm}
}
~\vskip 9mm
\caption{
As in Fig.~1 but with additional mass shift
$\Delta m_\rho = 150$ MeV.
}
\label{fig.2}
\end{figure}

\begin{figure} 
\centering{\epsfig{file=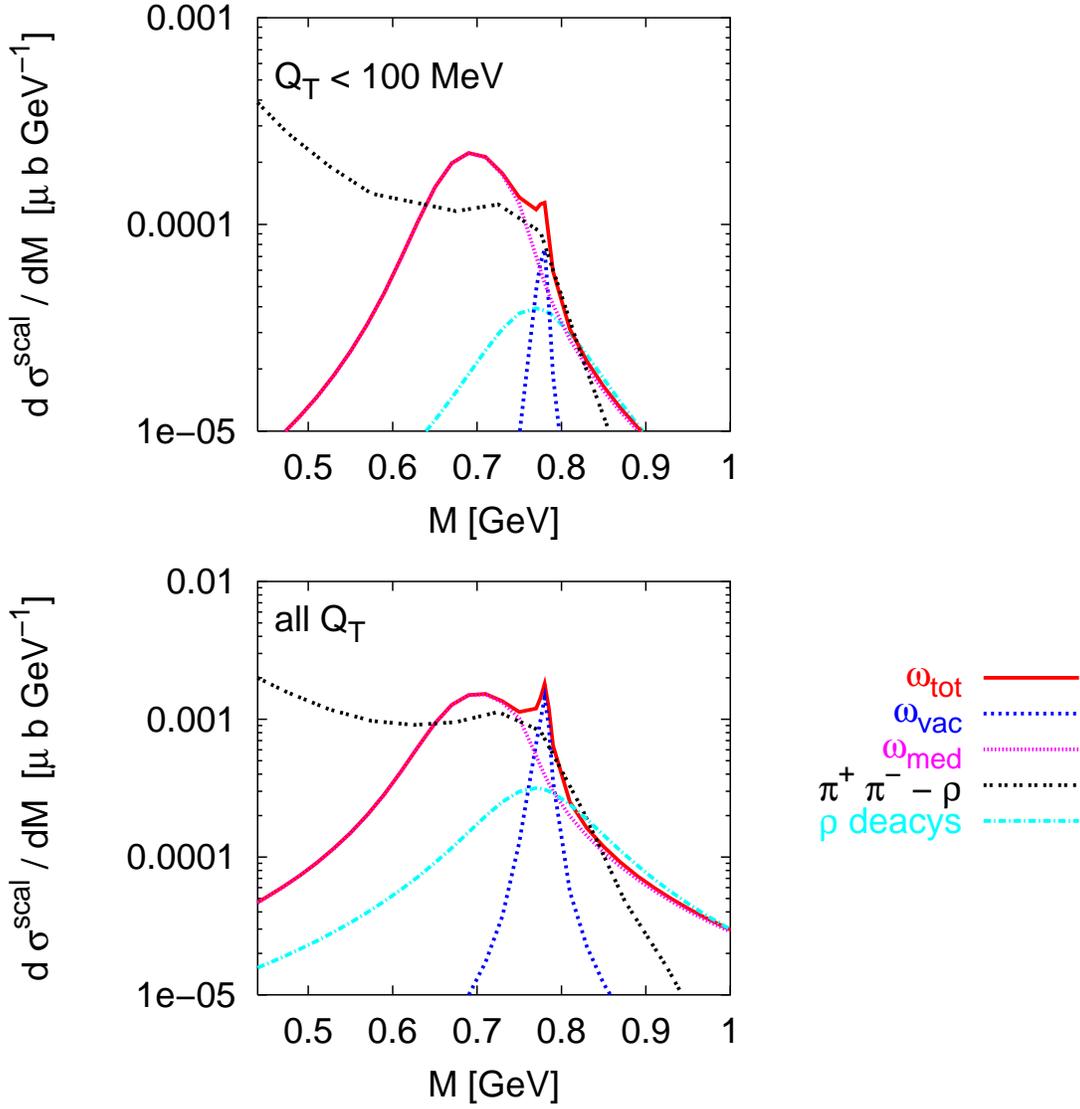,height=15cm}}
~\vskip 9mm
\caption{
Contributions to the di-electron spectrum 
$d \sigma^{scal} / d M$ for various bins of $Q_\perp$
for central collisions Au(1 AGeV) + Au.
The parameters are
$\delta \Gamma_\omega = 50$ MeV and
$\delta m_\omega = 70$ MeV.
For an orientation, also the $\rho$ contributions are shown,
but with vacuum parameters.
}
\label{fig.3}
\end{figure}

\begin{figure} 
\centering{\epsfig{file=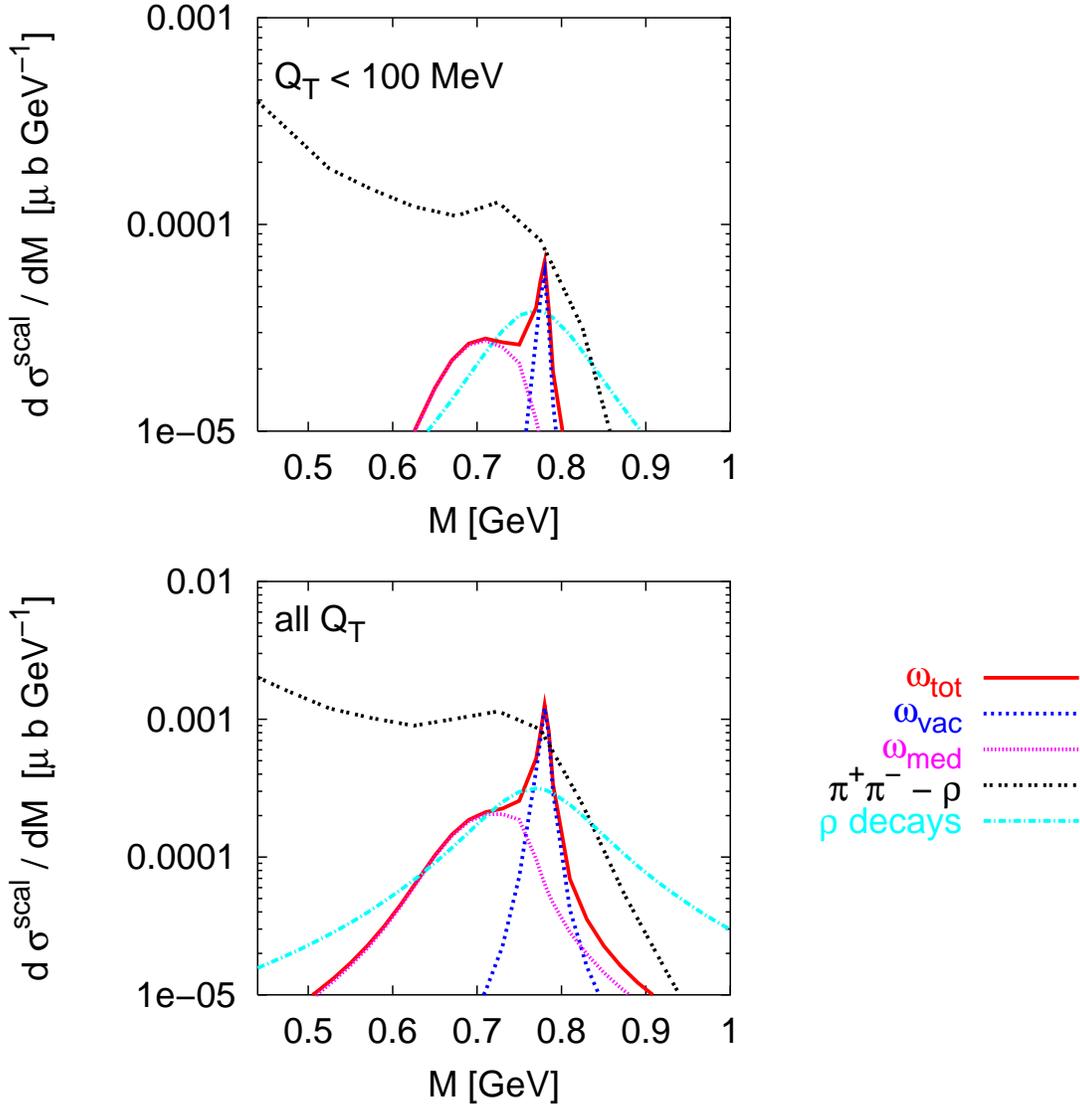,height=15cm}}
~\vskip 9mm
\caption{
As in Fig.~3, but with an $\omega$ branching ratio
as described in text.
}
\label{fig.4}
\end{figure}

\end{document}